\documentclass{optica-article}

\journal{opticajournal}
\articletype{Research Article}

\usepackage{booktabs}
\usepackage[numbers, square, sort&compress, sort]{natbib}
\usepackage{multirow}
\usepackage{lineno}
\usepackage{siunitx}

\renewcommand\thesubsection{\Alph{subsection}}

\begin{document}

\title{Beyond-Third-Order Quantum Coherence in Two-Dimensional Spectroscopy via Order-Selective Isolation}

\author{Xue Zhang,\authormark{1,$\dagger$} De-Ran Zhang,\authormark{2,$\dagger$} and Hui Dong\authormark{3,*}}

\address{\authormark{1}Graduate School of China Academy of Engineering Physics, No. 10 Xibeiwang East Road, Haidian District, Beijing, 100193, China\\
\authormark{2}Graduate School of China Academy of Engineering Physics, No. 10 Xibeiwang East Road, Haidian District, Beijing, 100193, China\\
\authormark{3}Graduate School of China Academy of Engineering Physics, No. 10 Xibeiwang East Road, Haidian District, Beijing, 100193, China\\
\authormark{$\dagger$}These authors contributed equally to this work.}

\email{\authormark{*}hdong@gscaep.ac.cn}


\begin{abstract*} 
A central challenge in nonlinear spectroscopy is the order-selective
readout of weak higher-order responses that spectrally overlap with
dominant lower-order signals. This bottleneck is particularly severe
in two-dimensional (2D) spectroscopy, where extending conventional
phase-cycling schemes to higher orders rapidly increases measurement
and analysis complexity. Here we introduce a computation-assisted
strategy that combines rotating-frame acquisition with a frame-shift tracking
algorithm to separate signals by their frame-dependent spectral shifts.
In a rubidium vapor experiment, we use this approach to isolate a
7th-order nonlinear contribution from coexisting 3rd-order components,
enabling direct access to higher-order quantum-coherence dynamics
without sacrificing operation at comparatively high pulse intensities.
The method is broadly compatible with multidimensional spectroscopy
platforms and provides a practical route to probing many-body and
collective ultrafast dynamics beyond third order.

\end{abstract*}

\section{Introduction}
 A longstanding challenge in nonlinear spectroscopy
is to increase the signal-to-noise ratio while retaining order-selective identification of dynamical  components. In the weak-field regime, linear
absorption and third-order pump--probe signals are often sufficient
to describe the response \cite{mukamel1995principles,cho2009two,hamm2011concepts}.
At higher excitation, however, additional nonlinear orders emerge
and overlap spectrally, obscuring the interpretation of pathways linked
to collective excitations and many-body interactions. This trade-off
between sensitivity and order specificity is a major bottleneck for
pushing multidimensional spectroscopy toward probing higher-order and many-body nonlinear responses.

Two-dimensional (2D) spectroscopy \cite{jonas2003two,Mukamel2000ARPC,tian2003femtosecond,Brixner2005,Myers2008,Shim2009,schlau2011two}
has enabled the isolation of third-order responses through
coherent-control strategies, non-collinear geometries separate pathways
via phase matching \cite{hybl2001two,jonas2003two,milota2013precise},
while collinear implementations use phase cycling to disentangle selected
quantum pathways \cite{tian2003femtosecond,tan2008theory,De2014}.
Yet, when higher-order contributions must be resolved or removed,
these approaches typically demand substantially larger phase-cycling
steps and heavier post-processing, leading to rapidly growing experimental
and computational complexity.

\begin{figure}
\begin{centering}
\includegraphics[width=8cm]{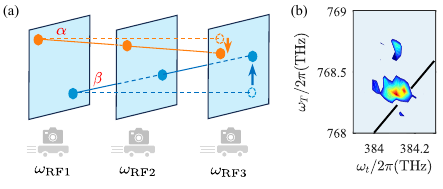}
\par\end{centering}
\caption{Scheme of the Frame-shift tracking separation algorithm for two-dimensional spectroscopy and the sorted 7-th signals.
(a) The scheme for the separation. Three spectra are sampled under different rotation frame frequency $\omega_{\mathrm{RF}}$, resembling the reduction of relative speed in a moving frame with constant velocity in classical mechanics. Movement of the spectral features (blue and orange points) is created with different slopes $k_{\alpha}=-1$ and $k_{\beta}=3$
for the frequency shift $\omega_{T}^{\mathrm{eff}}$ of the 3rd- and
7th-order signals, respectively. (b) The experimentally sorted 7th-order signals for rubidium vapor. The signal is sorted out by tracking the movement of the orange peak in subfigure (a) with the frame-shift tracking algorithm. \label{fig:mouse}}
\end{figure}

Recently, computation-assisted methods have emerged as a powerful
alternative to this limitation \cite{Kira2011,maly2023separating,krich2025separating}.
These methods introduce controlled modulations of experimental parameters
that generate order-dependent responses, followed by algorithmic separation.
For example, intensity cycling in pump--probe measurements exploits
nonlinear power scaling to allow disentangling mixed signal components \cite{maly2023separating}.
Such strategies relax intensity constraints while preserving interpretability,
and therefore offer a promising route to high-order multidimensional
spectroscopy.

Building on this computation-assisted framework, we combine
rotating-frame 2D measurements with a frame-shift tracking analysis that
extracts order-dependent peak-shift slopes across multiple rotating
frequencies. We experimentally validate the method in a rubidium
vapor cell \cite{Gao2016OptLett,Lomsadze2018}. As shown in Fig.~\ref{fig:mouse}(a),
the rotating-frame frequency $\omega_{\mathrm{RF}}$ induces systematic
displacements in the effective spectral coordinate $\omega_{T}^{\mathrm{eff}}$.
Tracking these displacements identifies distinct slope classes (lines
$\alpha$ and $\beta$), which directly encode nonlinear order. Using
this criterion, we isolate a seventh-order contribution [Fig.~\ref{fig:mouse}(b)]
from overlapping lower-order backgrounds. Importantly, the separation
remains effective at pulse intensities where higher-order signals
are experimentally accessible, enabling practical access to beyond-third-order
coherence dynamics.  

\section{Theory for Designing Computation-Assisted Spectroscopy}

\subsection{Pump–Probe Configuration and Measured Signal Formulation}
\begin{figure}[!t]
\centering
\includegraphics[width=7.5cm]{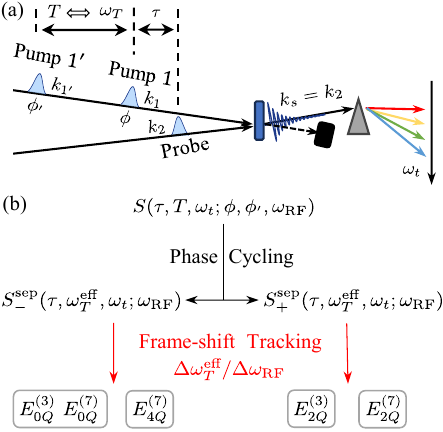}
\caption{Pulse sequences and the procedure for extracting higher-order signals.
(a) Pulse sequences of 2D spectroscopy under pump--probe geometry.
The wave vectors of the two pump pulses are denoted as $\mathbf{k}_{1}$
and $\mathbf{k}_{1'}$, with $\mathbf{k}_{1}=\mathbf{k}_{1'}$. The
wave vector of the probe pulse is $\mathbf{k}_{2}$. Additionally,
the probe pulse also acts as the local oscillator (LO), allowing for
heterodyned detection of signals with wave vector $\mathbf{k}_{s}=\mathbf{k}_{2}$.
Time zero is defined as the moment when the signals are generated.
(b) Signals $S(\tau,T,\omega_{t};\phi,\phi',\omega_{\text{RF}})$
are measured in three rotating frames with the same phase cycles.
In each frame, the signals are processed using phasing cycling and
causality enforcement to obtain the separated spectral signals $S_{+}^{\mathrm{sep}}(\tau,\omega_{T}^{\mathrm{eff}},\omega_{t};\omega_{\text{RF}})$
and $S_{-}^{\mathrm{sep}}(\tau,\omega_{T}^{\mathrm{eff}},\omega_{t};\omega_{\text{RF}})$.
The peak positions in each frame are determined via two-dimensional
Gaussian fitting. Signals of different orders are then separated based
on the distinct ratio $\Delta\omega_{T}^{\mathrm{eff}}/\Delta\omega_{\text{RF}}$
of their frame-dependent peak shifts, a procedure termed \textquotedblleft frame-shift
tracking\textquotedblright . The labels $E_{2Q}^{(3)},E_{0Q}^{(3)},E_{4Q}^{(7)},E_{2Q}^{(7)}$
and $E_{0Q}^{(7)}$ denote the signal fields that satisfy the phase-matching
conditions $\mathbf{k}_{s}=\mathbf{k}_{2}+\mathbf{k}_{1}-\mathbf{k}_{1'}$,
$\mathbf{k}_{2}-\mathbf{k}_{1}+\mathbf{k}_{1'}$, $\mathbf{k}_{2}+3\mathbf{k}_{1}-3\mathbf{k}_{1'}$,
$\mathbf{k}_{2}-3\mathbf{k}_{1}+3\mathbf{k}_{1'}$, and $\mathbf{k}_{2}+\mathbf{k}_{1}-2\mathbf{k}_{1}-\mathbf{k}_{1'}+2\mathbf{k}_{1'}$,
respectively. Certain signals, such as the field $E_{0Q}^{(7)}$,
share the same spectral frame and scaling ratio $\Delta\omega_{T}^{\mathrm{eff}}/\Delta\omega_{\text{RF}}$
as the 3rd-order $E_{0Q}^{(3)}$ signal. Yet, a subset of contributions
from $E_{0Q}^{(7)}$ is located at distinctly separate frequency positions,
revealing a more complex underlying structure. A detailed analysis
of these features is provided in Section 2.D. \label{fig:Experimental-arrangement.}}
\end{figure}

Our current setup for
the two-dimensional spectroscopy is under the pump-probe geometry,
where two pump pulses, generated by a pulse shaper, are collinear
with their wave vectors as $\mathbf{k}_{1}$ and $\mathbf{k}_{1'}$
($\mathbf{k}_{1'}=\mathbf{k}_{1}$), and the probe pulse propagates
in a non-collinear direction $\mathbf{k}_{2}$. Here we focus on a
modified pulse sequence where the probe pulse is designed to arrive
first at the sample with the time intervals between the three pulses
denoted as $\tau$ and $\ensuremath{T}$, as illustrated in Fig.
\ref{fig:Experimental-arrangement.}(a). The detection is along the
direction of the probe pulse with signal $E_{s}(\tau,T,\omega_{t})$,
where $\omega_{t}$ is the frequency of the signal, obtained directly
via a spectrometer. Such a sequence allows the investigation of double-quantum
(2Q) coherence \cite{Yang2008PRL,Dai2012two,Gao2016OptLett,Cai2024,timmer2024phase}
between the doubly excited state and the ground state within the time
interval $T$ .

In this sequence, the probe pulse also serves as the local oscillator
(LO). The signal is emitted only upon the arrival of the second pump
pulse. The signal field lags behind the probe pulse
by a time interval of $\tau+T$, resulting in an additional phase
accumulation $\exp[-i\omega_{t}(\tau+T)]$ during the propagation
of the probe pulse. The measured signal consequently takes the form,
\begin{eqnarray}
 &  & S(\tau,T,\omega_{t};\phi,\phi',\omega_{\mathrm{RF}})\nonumber \\
 & \approx & 2\text{Re}[E_{s}(\tau,T,\omega_{t};\phi,\phi',\omega_{\mathrm{RF}})E_{2}^{*}(\omega_{t})e^{i\omega_{t}(\tau+T)}],\label{eq:signal}
\end{eqnarray}
where $E_{2}(\omega_{t})$ represents the field of the probe pulse,
and $\phi$ ($\phi'$) denotes the phase of the first (second) pump
pulse. The multi-quantum 2D spectroscopy $S(\tau,\omega_{T},\omega_{t};\phi,\phi',\omega_{\mathrm{RF}})$
is obtained by sampling and Fourier transform of $S(\tau,T,\omega_{t};\phi,\phi',\omega_{\mathrm{RF}})$
with respect to the time variable $T$.In the signal-processing procedure of the double-quantum spectroscopy framework, a causality enforcement condition is additionally applied to preserve the causal ordering of the emitted signal field, which is a standard treatment in multidimensional coherent spectroscopy \cite{Myers2008,Cai2024}.

The present experimental design employs relatively strong excitation
pulses, giving rise to signals that contain significant contributions
beyond the typical third order. The signal $E_{s}(\tau,T,$\\
$\omega_{t};\phi,\phi',\omega_{\mathrm{RF}})$
along the probe direction ( i.e., $\mathbf{k}_{s}=\mathbf{k}_{2}$)
can be decomposed into different components as $E_{s}(\tau,T,\omega_{t};\phi,\phi',\omega_{\mathrm{RF}})=\sum_{n}E_{s}^{(2n+1)}(\tau,T,\omega_{t};\phi,\phi',\omega_{\mathrm{RF}})$
of the $(2n+1)$th order as $\mathbf{k}_{s}=\mathbf{k}_{2}+n_{1}\mathbf{k}_{1}-m_{1}\mathbf{k}_{1}+n_{2}\mathbf{k}_{1'}-m_{2}\mathbf{k}_{1'}$,
where $n_{1}+m_{1}+n_{2}+m_{2}=2n$, $n=1,2,3,\cdots$ with $n_{1}-m_{1}+n_{2}-m_{2}=0$. We define the configurations satisfying $(n_{1},m_{1},n_{2},m_{2})=(n,0,0,n)$ or $(0,n,n,0)$ as fully symmetric configurations, while all remaining cases are referred to as asymmetric configurations.

These $(2n+1)$th-order components result from different combinations
of $2n$ interactions with the two collinear pump pulses $\mathbf{k}_{1}$
and $\mathbf{k}_{1'}$, while the probe pulse is assumed to be sufficiently
weak to interact only once with the sample. In the excitation situations
governed by the phase-matching condition, the sign of a wave vector
is directly related to the change of excitation numbers in the density
matrix element $\rho_{\alpha,\alpha'}$, where $\left|\alpha\right\rangle $
and $\left|\alpha'\right\rangle $ are the eigen-states of the system
Hamiltonian. A positive wave vector (e.g., $+\mathbf{k}_{1}$) increases
the excitation number $\alpha$ or decreases $\alpha\prime$, while
a negative wave vector (e.g., $-\mathbf{k}_{1}$) decreases $\alpha$
or increases $\alpha\prime$. The measured signal arises from different
$M$-th quantum coherences within the time interval $T$, which correspond
to density matrix elements $\rho_{n_{1}+1,m_{1}}$ with the coherence
order $M=\left|n_{1}+1-m_{1}\right|$, indicating the absolute difference
in the number of excitations between the two states $\left|n_{1}+1\right\rangle $
and $\left|m\right\rangle $. Critically, under strong pump fields,
a given multi-quantum coherence signal (e.g., 2Q) during $T$ comprises
contributions from multiple nonlinear orders.

\subsection{Fully Symmetric Configuration and Coherence Separation Strategy}

In this letter, we first and primarily focus on two types of fully symmetric $(2n+1)$th-order signals,
(i) with the phase-matching condition $\mathbf{k}_{s}=\mathbf{k}_{2}+n\mathbf{k}_{1}-n\mathbf{k}_{1'}$
i.e., $n_{1}=m_{2}=n$ and $m_{1}=n_{2}=0$, which induces the highest-order
coherence $\rho_{n+1,0}$ with $M=n+1$, i.e., and (ii) with $\mathbf{k}_{s}=\mathbf{k}_{2}-n\mathbf{k}_{1}+n\mathbf{k}_{1'}$
($n_{1}=m_{2}=0$ and $m_{1}=n_{2}=n$) which results in the second-highest
coherence $\rho_{1,n}$ and $\rho_{0,n-1}$ with $M=n-1$, during
the delay time $T$. The corresponding signal fields are denoted as
$E_{(n+1)Q}^{(2n+1)}(\tau,T,\omega_{t};\phi,\phi',\omega_{\mathrm{RF}})$
and $E_{(n-1)Q}^{(2n+1)}(\tau,T,\omega_{t};\phi,\phi',\omega_{\mathrm{RF}})$,
respectively. We note here that any signal with coherence on the order
$M=n$ is cycled out in the subsequent signal processing via phase
cycling. Based on this fully symmetric configuration, which supports the generation of both the highest-order and second-highest-order coherences, we develop the corresponding separation scheme. The remaining configurations are discussed in detail in the following section, where we take the seventh-order signal as a representative example for a comprehensive analysis.

The overall signal separation proceeds through two key steps: First,
phase cycling separates the acquired signal into two distinct spectra.
Then, by combining a rotating-frame scheme with a frame-shift tracking algorithm,
high-order signal components are identified and isolated, as illustrated
in Fig. 2(b). We now elaborate on these two core procedures as follows.

In our partially collinear geometry, $(n-1)Q$ and $(n+1)Q$ signal
fields exhibit phase dependencies,
\begin{equation}
E_{(n+1)Q}^{(2n+1)}(\tau,T,\omega_{t};\phi,\phi',\omega_{\mathrm{RF}})\propto\text{exp\ensuremath{[in(\phi-\phi')]}},\label{eq:n+1}
\end{equation}
\begin{equation}
E_{(n-1)Q}^{(2n+1)}(\tau,T,\omega_{t};\phi,\phi',\omega_{\mathrm{RF}})\propto\text{exp\ensuremath{[in(-\phi+\phi')]}.}\label{eq:n-1}
\end{equation}
In our protocol, we implement a four-step phase cycle: $\Phi_{1}=(\phi=0,\phi'=0)$,
$\Phi_{2}=(\phi=0,\phi'=\pi/2)$, $\Phi_{3}=(\phi=0,\phi'=\pi)$,and
$\Phi_{4}=(\phi=0,\phi'=3\pi/2)$. The measured signals are combined
into $S_{+}^{\mathrm{sep}}(\tau,T,\omega_{t};\omega_{\mathrm{RF}})=S(\tau,T,\omega_{t};0,0,\omega_{\mathrm{RF}})+iS(\tau,T,\omega_{t};0,\pi/2,\omega_{\mathrm{RF}})$
and $S_{-}^{\mathrm{sep}}(\tau,T,\omega_{t};\omega_{\mathrm{RF}})=S(\tau,T,\omega_{t};0,0,\omega_{\mathrm{RF}})-iS(\tau,T,\omega_{t};0,\pi/2,\omega_{\mathrm{RF}})$,
where $S(\tau,T,\omega_{t};0,0,$\\
$\omega_{\mathrm{RF}})=S^{0}(\tau,T,\omega_{t};0,0,\omega_{\mathrm{RF}})-S^{0}(\tau,T,\omega_{t};0,\pi,\omega_{\mathrm{RF}})$,
and $S(\tau,T,\omega_{t};0,\pi/2,\omega_{\mathrm{RF}})=S^{0}(\tau,T,$\\
$\omega_{t};0,\pi/2,\omega_{\mathrm{RF}})-S^{0}(\tau,T,\omega_{t};0,3\pi/2,\omega_{\mathrm{RF}})$,
with $S^{0}$ denoting the raw experimentally acquired signals. Such
phase cycling eliminates transient absorption signal \cite{Shim2009},
and also helps to isolate different coherences \cite{Myers2008}
as follows. Signals with even $n$ (e.g., fifth-order with $n=2$)
are completely suppressed, while signals with odd $n$ are separated
into $S_{+}^{\mathrm{sep}}(\tau,T,\omega_{t};\omega_{\mathrm{RF}})$
and $S_{-}^{\mathrm{sep}}(\tau,T,\omega_{t};\omega_{\mathrm{RF}})$
depending on $n$: for $n=1,5,9,\cdots$, the $(n+1)Q$ contributions
appear in $S_{+}^{\mathrm{sep}}(\tau,T,\omega_{t};\omega_{\mathrm{RF}})$,
whereas the $(n-1)Q$ contributions appear in $S_{-}^{\mathrm{sep}}(\tau,T,\omega_{t};\omega_{\mathrm{RF}})$;
for $n=3,7,11,\cdots$, the $(n-1)Q$ contributions appear in $S_{+}^{\mathrm{sep}}(\tau,T,\omega_{t};\omega_{\mathrm{RF}})$,
whereas the $(n+1)Q$ contributions appear in $S_{-}^{\mathrm{sep}}(\tau,T,\omega_{t};\omega_{\mathrm{RF}})$.
For example, the $E_{2Q}^{(7)}$ and $E_{2Q}^{(3)}$ are isolated
into $S_{+}^{\mathrm{sep}}(\tau,T,\omega_{t};\omega_{\mathrm{RF}})$,
while the $E_{4Q}^{(7)}$ and $E_{0Q}^{(3)}$ are isolated into $S_{-}^{\mathrm{sep}}(\tau,T,\omega_{t};\omega_{\mathrm{RF}})$.

Next, we develop a new method based on the rotating frame to further
isolate different signals in $S_{+}^{\mathrm{sep}}(\tau,T,\omega_{t};\omega_{\mathrm{RF}})$
and $S_{-}^{\mathrm{sep}}(\tau,T,\omega_{t};\omega_{\mathrm{RF}})$
spectra. To reduce the sampling rate, a rotating frame with frequency
$\omega_{\mathrm{RF}}$ is applied to the first pump pulse via a pulse
shaper, which modifies the temporal oscillations of the $(2n+1)$th-order
signal fields during $T$ as follows:
\begin{eqnarray}
E_{(n+1)Q}^{(2n+1)}(\tau,T,\omega_{t};\omega_{\mathrm{RF}}) & \propto & \exp[-i(\omega_{n+1,0}-n\omega_{\mathrm{RF}})T],
\end{eqnarray}
and
\begin{eqnarray}
E_{(n-1)Q}^{(2n+1)}(\tau,T,\omega_{t};\omega_{\mathrm{RF}}) & \propto & \exp[-i(-\omega_{n,1}+n\omega_{\mathrm{RF}})T]\\
 & \mathrm{or} & \exp[-i(-\omega_{n-1,0}+n\omega_{\mathrm{RF}})T],\nonumber 
\end{eqnarray}
where $\omega_{\alpha,\alpha'}\equiv\omega_{\alpha}-\omega_{\alpha'}$
is the energy gap between the energy levels with $\alpha$ and $\alpha'$
of the density matrix $\rho_{\alpha,\alpha'}$. Incorporating the
propagation phase factor $\exp[-i\omega_{t}(\tau+T)]$ in Eq.~(\ref{eq:signal}),
we derive the effective oscillation frequency $\omega_{T}^{\text{eff}}$
that appear in the Fourier transform spectra $S_{\pm}^{\mathrm{sep}}(\tau,\omega_{T}^{\text{eff}},\omega_{t};\omega_{\mathrm{RF}})$
as $\omega_{T}^{\text{eff}}=\omega_{n+1,0}-n\omega_{\mathrm{RF}}-\omega_{t}$
for the $(n+1)Q$ signals and $\omega_{T}^{\text{eff}}=-\omega_{n,1}+n\omega_{\mathrm{RF}}-\omega_{t}$
for the $(n-1)Q$ signals. The shift rate of spectral peaks along
the $\omega_{T}^{\text{eff}}$ axis with respect to $\omega_{\mathrm{RF}}$,
$\Delta\omega_{T}^{\text{eff}}/\Delta\omega_{\mathrm{RF}}$, provides
an intrinsic signature of the nonlinear order. Such a rate difference
$\Delta\omega_{T}^{\text{eff}}/\Delta\omega_{\mathrm{RF}}=\pm n$
between the $(n+1)Q$ and $(n-1)Q$ signal allows for both the separation
and unambiguous identification of signal components.

\subsection{General Behavior of High-Order Pathways: A Seventh-Order Analysis}

To comprehensively model the isolation of high-order signals, we must
account for all nonlinear processes that induce signals along the
phase-matching direction $\mathbf{k}_{s}=\mathbf{k}_{2}$. We therefore
adopt the general expression for the signal wave vectors as defined
earlier: $\mathbf{k}_{s}=\mathbf{k}_{2}+n_{1}\mathbf{k}_{1}-m_{1}\mathbf{k}_{1}+n_{2}\mathbf{k}_{1'}-m_{2}\mathbf{k}_{1'}$,
where $\mathbf{k}_{1}=\mathbf{k}_{1'}$. Under this model, the phase-matching
condition is satisfied with $n_{1}+n_{2}=m_{1}+m_{2}$.

To illustrate all the possible pathways related to the current discussion,
we divide the pathways for the signals on the $(2n+1)$th-order into
two scenarios, (1) $n_{1}+m_{1}=n_{2}+m_{2}=n$; (2) $n_{1}+m_{1}\neq n_{2}+m_{2}$.
To clearly demonstrate the properties of these pathways, we take an
example for the 7-th order with $n=3$. We mainly focus on the resulting
coherence types, which relate to the frequency $\omega_{T}$.

\begin{table*}[!t]
\begin{center}
\caption{Seventh-order signal components and their characteristics.}
\centering
\small
\setlength{\tabcolsep}{1pt}
\resizebox{0.98\textwidth}{!}{%
\begin{tabular}{r@{\extracolsep{0pt}}lccccr@{\extracolsep{0pt}}lr}
\toprule 
\multicolumn{2}{c}{Situation} & Signal Direction $\mathbf{k}_{s}$ & Coherence Type & Effective Frequency $\omega_{T}^{\text{eff}}$ & $\frac{\Delta\omega_{T}^{\text{eff}}}{\Delta\omega_{\mathrm{RF}}}$ & \multicolumn{2}{c}{Assigned Spectrum} & Separation Scheme \\
\midrule 
\multicolumn{2}{c}{1} & $\mathbf{k}_{2}+3\mathbf{k}_{1}-3\mathbf{k}_{1'}$ & $\rho_{4,0}$(4Q) & $\omega_{4,0}-3\omega_{\mathrm{RF}}-\omega_{t}$ & -3 & \multicolumn{2}{c}{$S_{-}^{\mathrm{sep}}$} & Velocity-resolved \\
\midrule 
\multicolumn{2}{c}{2} & \multirow{2}{*}[-1pt]{$\mathbf{k}_{2}-3\mathbf{k}_{1}+3\mathbf{k}_{1'}$} & $\rho_{3,1}$(2Q) & $-\omega_{3,1}+3\omega_{\mathrm{RF}}-\omega_{t}$ & +3 & \multicolumn{2}{c}{\multirow{2}{*}{$S_{+}^{\mathrm{sep}}$}} & \multirow{2}{*}{Velocity-resolved} \\
\multicolumn{2}{c}{3} & & $\rho_{2,0}$(2Q) & $-\omega_{2,0}+3\omega_{\mathrm{RF}}-\omega_{t}$ & +3 & & \\
\midrule 
\multicolumn{2}{c}{4} & \multirow{2}{*}[-1pt]{$\mathbf{k}_{2}+2\mathbf{k}_{1}-\mathbf{k}_{1}-2\mathbf{k}_{1'}+\mathbf{k}_{1'}$} & $\rho_{3,1}$(2Q) & $\omega_{3,1}-\omega_{\mathrm{RF}}-\omega_{t}$ & -1 & \multicolumn{2}{c}{\multirow{2}{*}{$S_{+}^{\mathrm{sep}}$}} & \multirow{2}{*}{Degenerate (with 3rd)} \\
\multicolumn{2}{c}{5} & & $\rho_{2,0}$(2Q) & $\omega_{2,0}-\omega_{\mathrm{RF}}-\omega_{t}$ & -1 & & \\
\midrule 
\multicolumn{2}{c}{6} & \multirow{3}{*}[-1pt]{$\mathbf{k}_{2}+\mathbf{k}_{1}-2\mathbf{k}_{1}-\mathbf{k}_{1'}+2\mathbf{k}_{1'}$} & $\rho_{2,2}$(0Q) & $-\omega_{2,2}+\omega_{\mathrm{RF}}-\omega_{t}$ & +1 & \multicolumn{2}{c}{\multirow{3}{*}{$S_{-}^{\mathrm{sep}}$}} & Frequency-resolved \\
\multicolumn{2}{c}{7} & & $\rho_{1,1}$(0Q) & $-\omega_{1,1}+\omega_{\mathrm{RF}}-\omega_{t}$ & +1 & & & Degenerate (with 3rd) \\
\multicolumn{2}{c}{8} & & $\rho_{0,0}$(0Q) & $-\omega_{0,0}+\omega_{\mathrm{RF}}-\omega_{t}$ & +1 & & & Degenerate (with 3rd) \\
\midrule 
\multicolumn{2}{c}{9} & $\mathbf{k}_{2}+\mathbf{k}_{1}-3\mathbf{k}_{1'}+2\mathbf{k}_{1'}$ & $\rho_{2,0}$(2Q) & $\omega_{2,0}-\omega_{\mathrm{RF}}-\omega_{t}$ & -1 & \multicolumn{2}{c}{$S_{+}^{\mathrm{sep}}$} & Degenerate (with 3rd) \\
\midrule 
\multicolumn{2}{c}{10} & \multirow{2}{*}[-1pt]{$\mathbf{k}_{2}-\mathbf{k}_{1}+3\mathbf{k}_{1'}-2\mathbf{k}_{1'}$} & $\rho_{1,1}$(0Q) & $-\omega_{1,1}+\omega_{\mathrm{RF}}-\omega_{t}$ & +1 & \multicolumn{2}{c}{\multirow{2}{*}{$S_{-}^{\mathrm{sep}}$}} & \multirow{2}{*}{Degenerate (with 3rd)} \\
\multicolumn{2}{c}{11} & & $\rho_{0,0}$(0Q) & $-\omega_{0,0}+\omega_{\mathrm{RF}}-\omega_{t}$ & +1 & & \\
\midrule 
\multicolumn{2}{c}{12} & \multirow{3}{*}[-1pt]{$\mathbf{k}_{2}+3\mathbf{k}_{1}-2\mathbf{k}_{1}-\mathbf{k}_{1'}$} & $\rho_{4,2}$(2Q) & $\omega_{4,2}-\omega_{\mathrm{RF}}-\omega_{t}$ & -1 & \multicolumn{2}{c}{\multirow{3}{*}{$S_{+}^{\mathrm{sep}}$}} & Frequency-resolved \\
\multicolumn{2}{c}{13} & & $\rho_{3,1}$(2Q) & $\omega_{3,1}-\omega_{\mathrm{RF}}-\omega_{t}$ & -1 & & & Frequency-resolved \\
\multicolumn{2}{c}{14} & & $\rho_{2,0}$(2Q) & $\omega_{2,0}-\omega_{\mathrm{RF}}-\omega_{t}$ & -1 & & & Degenerate (with 3rd) \\
\midrule 
\multicolumn{2}{c}{15} & \multirow{4}{*}[-1pt]{$\mathbf{k}_{2}+2\mathbf{k}_{1}-3\mathbf{k}_{1}+\mathbf{k}_{1'}$} & $\rho_{3,3}$(0Q) & $-\omega_{3,3}+\omega_{\mathrm{RF}}-\omega_{t}$ & +1 & \multicolumn{2}{c}{\multirow{4}{*}{$S_{-}^{\mathrm{sep}}$}} & Frequency-resolved \\
\multicolumn{2}{c}{16} & & $\rho_{2,2}$(0Q) & $-\omega_{2,2}+\omega_{\mathrm{RF}}-\omega_{t}$ & +1 & & & \parbox{3cm}{\centering Degenerate (with 3rd \\ or case 6)} \\
\multicolumn{2}{c}{17} & & $\rho_{1,1}$(0Q) & $-\omega_{1,1}+\omega_{\mathrm{RF}}-\omega_{t}$ & +1 & & & Degenerate (with 3rd) \\
\multicolumn{2}{c}{18} & & $\rho_{0,0}$(0Q) & $-\omega_{0,0}+\omega_{\mathrm{RF}}-\omega_{t}$ & +1 & & & Degenerate (with 3rd) \\
\midrule 
\multicolumn{2}{c}{19} & $\mathbf{k}_{2}+3\mathbf{k}_{1}-\mathbf{k}_{1}-2\mathbf{k}_{1'}$, $\mathbf{k}_{2}+2\mathbf{k}_{1}-3\mathbf{k}_{1'}+\mathbf{k}_{1'}$ & 3Q & --- & --- & \multicolumn{2}{c}{---} & Phase-cycled out \\
\midrule 
\multicolumn{2}{c}{20} & $\mathbf{k}_{2}-3\mathbf{k}_{1}+\mathbf{k}_{1}+2\mathbf{k}_{1'}$, $\mathbf{k}_{2}-2\mathbf{k}_{1}+3\mathbf{k}_{1'}-\mathbf{k}_{1'}$ & 1Q & --- & --- & \multicolumn{2}{c}{---} & Phase-cycled out \\
\midrule 
\multicolumn{2}{c}{21} & $\mathbf{k}_{2}+2\mathbf{k}_{1}-2\mathbf{k}_{1}+\mathbf{k}_{1'}-\mathbf{k}_{1'}$, $\mathbf{k}_{2}+\mathbf{k}_{1}-\mathbf{k}_{1}+2\mathbf{k}_{1'}-2\mathbf{k}_{1'}$ & 1Q & --- & --- & \multicolumn{2}{c}{---} & Phase-cycled out \\
\bottomrule
\end{tabular}
}
\label{table1_em}
\end{center}
\end{table*}

\textbf{Scenario (1) with $n_{1}+m_{1}=n_{2}+m_{2}=3$}. In this scenario,
the first case is $(n_{1}=m_{2}=3$, $m_{1}=n_{2}=0)$ and $(m_{1}=n_{2}=3$,
$n_{1}=m_{2}=0)$, corresponding to the phase matching conditions
$\mathbf{k}_{s}=\mathbf{k}_{2}+3\mathbf{k}_{1}-3\mathbf{k}_{1'}$
and $\mathbf{k}_{s}=\mathbf{k}_{2}-3\mathbf{k}_{1}+3\mathbf{k}_{1'}$.
These pathways belong to the fully symmetric excitation configuration introduced in the previous subsection, and presented as lines
1 - 3 in the Table. \ref{table1_em}. As discussed in the previous subsection, the fully symmetric configuration constitutes the central design principle of our separation framework, since it directly generates the highest- and second-highest-order coherences with well-defined rotating-frame shift rates. For the second phase-matching condition, a distinct coherence
$\rho_{3,1}$ emerges with oscillation frequency $\omega_{3,1}$. This frequency arises
from multi-atom interactions and corresponds to a distinct spectral
position, as indicated in line 2 of the Table. \ref{table1_em}. However,
due to potentially weak interaction strengths or minute energy-level
shifts, the corresponding signal may be too weak to be discernible
in our spectra. We note this possibility for reference, though it
does not manifest as a distinct feature in our current measurements.In the following discussion, we further extend the analysis to asymmetric excitation configurations and examine their corresponding spectral behaviors within the same theoretical framework.

The second case includes two phase matching conditions, $\mathbf{k}_{s}=\mathbf{k}_{2}+2\mathbf{k}_{1}-\mathbf{k}_{1}+\mathbf{k}_{1'}-2\mathbf{k}_{1'}$
and $\mathbf{k}_{s}=\mathbf{k}_{2}+\mathbf{k}_{1}-2\mathbf{k}_{1}+2\mathbf{k}_{1'}-1\mathbf{k}_{1'}$.
For the first phase matching condition, the second pulse, interacting
with the system three times, results in the coherence $\rho_{3,1}$
or $\rho_{2,0}$ during the waiting time $T$. For the signal related
to coherence $\rho_{2,0}$ (line 5 in Table. \ref{table1_em}), the
oscillation frequency $\omega_{T}=\omega_{2,0}$ is exactly the same
as that created by the third order response along the phase matching
condition $\mathbf{k}_{s}=\mathbf{k}_{2}+\mathbf{k}_{1}-\mathbf{k}_{1'}$.
Such a signal cannot be separated with the current method, yet may be distinguished
with the power cycling in Ref. \cite{maly2023separating}. We also
would like to remark that the isolation of these types of high-order
signals will not provide any additional information due to the exact
dynamics. And we name this type of high-order signal as the ``shielding''
effect on the low-order signal. The signal related to the coherence
$\rho_{3,1}$ (line 4 in Table. \ref{table1_em}) may have a different
oscillation frequency $\omega_{T}=\omega_{3,1}$, which may be different
from the interaction between Rb atoms. Such a difference shall result
in a different position on the 2D spectra. The second phase matching
condition may also give rise to a different coherence $\rho_{f,D_{2}+D_{2}}$
(line 6 in Table. \ref{table1_em}), where the state $\left|f\right\rangle $
is the $5^{2}S_{1/2}\leftrightarrow5^{2}D$ transition state created
by the probe and pump pulse, and the state $\left|D_{2}+D_{2}\right\rangle $
denotes a double $D_{2}$ transition.

These observation lead to a general rule for higher-order signals,
whenever the pump-probe pulses generate higher-oder signals with the
same dynamics as lower-order signal, these signals cannot be distinguished
by the current methods. These shielding signals may not provide additional
information, and shall not require extra effort to isolate them.

\textbf{Scenario (2) with $n_{1}+m_{1}\neq n_{2}+m_{2}$}. This scenario
includes four cases: ($n_{1}+m_{1}=1,n_{2}+m_{2}=5$), ($n_{1}+m_{1}=2,n_{2}+m_{2}=4$),
($n_{1}+m_{1}=4,n_{2}+m_{2}=2$), and ($n_{1}+m_{1}=5,n_{2}+m_{2}=1$).
The cases ($n_{1}+m_{1}=2,n_{2}+m_{2}=4$) and ($n_{1}+m_{1}=4,n_{2}+m_{2}=2$)
resemble the even $n$ situation and are
canceled by phase-cycling procedure; they correspond to lines 19,
20, and 21 in Table. \ref{table1_em}, respectively.

For the remaining cases ($n_{1}+m_{1}=1,n_{2}+m_{2}=5$) and ($n_{1}+m_{1}=5,n_{2}+m_{2}=1$),
the first case ($n_{1}+m_{1}=1,n_{2}+m_{2}=5$) includes two phase
matching conditions, $\mathbf{k}_{s}=\mathbf{k}_{2}+\mathbf{k}_{1}-3\mathbf{k}_{1'}+2\mathbf{k}_{1'}$
and $\mathbf{k}_{s}=\mathbf{k}_{2}-\mathbf{k}_{1}+3\mathbf{k}_{1'}-2\mathbf{k}_{1'}$,
in which the second pulse interacts once with the system. For the
first phase matching condition, the second pulse generates the coherence
$\rho_{2,0}$ (line 9 in Table. \ref{table1_em}) during the waiting
time $T$, oscillating at $\omega_{T}=\omega_{2,0}$. This frequency
is identical to that produced by the 3rd-order signal with $\mathbf{k}_{s}=\mathbf{k}_{2}+\mathbf{k}_{1}-\mathbf{k}_{1'}$,
and therefore cannot be separated. For the second phase matching condition,
the coherence is either $\rho_{1,1}$ (line 10 in Table. \ref{table1_em})
or $\rho_{0,0}$ (line 11 in Table. \ref{table1_em}), oscillating
at $\omega_{1,1}$ or $\omega_{0,0}$, which likewise coincides with
the third order signal along $\mathbf{k}_{s}=\mathbf{k}_{2}-\mathbf{k}_{1}+\mathbf{k}_{1'}$;
hence it also cannot be separated.

The second case ($n_{1}+m_{1}=5,n_{2}+m_{2}=1$) also includes two
phase matching conditions, $\mathbf{k}_{s}=\mathbf{k}_{2}+3\mathbf{k}_{1}-2\mathbf{k}_{1}-\mathbf{k}_{1'}$
and $\mathbf{k}_{s}=\mathbf{k}_{2}+2\mathbf{k}_{1}-3\mathbf{k}_{1}+\mathbf{k}_{1'}$,
where the second pulse interacts with the system for five times. For
the first phase matching condition, the second pulse yields coherence
$\rho_{4,2}$ (line 12 in Table. \ref{table1_em}), $\rho_{3,1}$
(line 13 in Table. \ref{table1_em}), or $\rho_{2,0}$ (line 14 in
Table. \ref{table1_em}), oscillating at $\omega_{4,2}$, $\omega_{3,1}$,
or $\omega_{2,0}$. The signals oscillating at $\omega_{4,2}$ and
$\omega_{3,1}$ arise from distinct atomic interactions and are thus
expected to appear at different positions in the 2D spectra. In contrast,
the component oscillating at $\omega_{T}=\omega_{2,0}$ is degenerate
with the third\nobreakdash-order signal along $\mathbf{k}_{s}=\mathbf{k}_{2}+\mathbf{k}_{1}-\mathbf{k}_{1'}$
and cannot be separated. For the second phase matching condition,
the second pulse yields coherence $\rho_{3,3}$ (line 15 in Table.
\ref{table1_em}) $\rho_{2,2}$ (line 16 in Table. \ref{table1_em}),
$\rho_{1,1}$ (line 17 in Table. \ref{table1_em}), or $\rho_{0,0}$
(line 18 in Table. \ref{table1_em}), oscillating at $\omega_{3,3}$,
$\omega_{2,2}$, $\omega_{1,1}$, or $\omega_{0,0}$.The frequency
oscillating at $\omega_{3,3}$ is also expected to be spectrally distinct.
The component at $\omega_{2,2}$ yields the same result as the line 6
in Table. \ref{table1_em} discussed above. Finally, the components
at $\omega_{1,1}$ and $\omega_{0,0}$ are degenerate with the third
order signal along $\mathbf{k}_{s}=\mathbf{k}_{2}-\mathbf{k}_{1}+\mathbf{k}_{1'}$
and thus cannot be separated with current methods.

To conclude, a vast majority of signals not related to $\mathbf{k}_{s}=\mathbf{k}_{2}\pm n\mathbf{k}_{1}\mp n\mathbf{k}_{1'}$,
degenerate into lower-order signals, resulting in complete spectral
overlap. This makes them intrinsically inseparable in our measurement.
The remaining few special situations are readily identifiable through
their dynamics, as exemplified by $\mathbf{k}_{s}=\mathbf{k}_{2}+\mathbf{k}_{1}-2\mathbf{k}_{1}+2\mathbf{k}_{1'}-1\mathbf{k}_{1'}$.

\subsection{Experimental Signal Extraction}

To extract spectral shifts across different orders, we employ a frame-shift tracking
algorithm, which is used in computer science to determine the position
changes by comparing subsequent frame during
the movement. Here, the movement of the spectra is created by changing
the frequency $\omega_{\mathrm{RF}}$ of the rotating frame. The procedure
of our frame-shift tracking for spectra begins by applying an absolute threshold
of 0.095 to the 2D spectra to suppress weak background signals. The
remaining features are clustered to isolate distinct peaks. Each peak
is then fitted with a two-dimensional Gaussian function to determine
its center coordinates $(\omega_{x},\omega_{y})$

\begin{equation}
S(\omega_{t},\omega_{T}^{\text{eff}})=a_{0}\exp[-\frac{(\omega_{t}-\omega_{x})^{2}}{\sigma_{x}^{2}}-\frac{(\omega_{T}^{\text{eff}}+\omega_{t}-\omega_{y})^{2}}{\sigma_{y}^{2}}]+b_{0},\label{eq:6}
\end{equation}
where $a_{0}$ is the amplitude, $\sigma_{x}$ and $\sigma_{y}$ are
the corresponding widths, and $b_{0}$ is a constant background. The
term $\omega_{T}^{\text{eff}}+\omega_{t}$ captures the tilt of the
signal induced by the phase factor $\exp[i\omega_{t}(\tau+T)]$ in
Eq.~(\ref{eq:signal}). 

To track the displacement of peak centers $(\Delta\omega_{x},\Delta\omega_{y})$
between two rotating frames, we employ the Hungarian algorithm\cite{Bewley2016,Sahbani2016},
which solves the optimal assignment between peak sets by minimizing
the total pairwise distance cost, similar to the algorithm of an optical
mouse. This ensures robust and consistent peak matching as signals
shift. The order of each signal is subsequently determined by comparing
the observed relative shift $\Delta\omega_{y}/\Delta\omega_{\mathrm{RF}}$
along the $\omega_{T}^{\mathrm{eff}}$ axis, as shown in Fig. \ref{fig:mouse}(a)
with the representative lines $\alpha$ and $\beta$. Further details
of the tracking and isolation are presented later along with the experimental
verification. We would remark that the current scheme with the rotating
frame relies on differences in phase evolution or dynamics during
the waiting time $T$. When signals of different orders are fully
degenerate in frequency and exhibit identical temporal evolution,
they remain indistinguishable within the present framework.

\section{Experimental verification}
We validate this approach experimentally
using a $^{87}\mathrm{Rb}$ vapor system contained in a borosilicate
cell filled with 450 Torr of nitrogen buffer gas to induce collisional
broadening. The cell is stabilized at \SI{120.0}{\degreeCelsius} (\textpm \SI{0.1}{\degreeCelsius}),
yielding a high-density thermal vapor (approximately $1.65\times 10^{13}\mathrm{cm}^{-3}$).

Pump and probe pulses are generated by a Ti:Sapphire amplifier laser system producing pulses (800 nm, 30.8 fs) at a 1 kHz repetition rate. The central wavelength is tuned using an optical parametric amplifier (OPA), which allows spectral shifting to 382.9THz (783nm) for the present experiment.

The pump–probe excitation scheme is implemented using a pulse-shaping system consisting of an acousto-optic modulator (AOM), two parabolic mirrors, and a pair of gratings in a 4-f configuration. This setup converts a single pump pulse into two phase-controlled collinear pump pulses with a variable temporal separation T. The delay $\tau$ between the probe pulse and the first pump pulse is controlled by a mechanical delay stage in the probe arm. Both pump and probe beams are focused onto the sample using a parabolic mirror to ensure optimal spatial overlap, and their polarizations are aligned horizontally using half-wave plates and polarizers prior to entering the sample.

\begin{figure}[t]
\centering
\centering{}\includegraphics{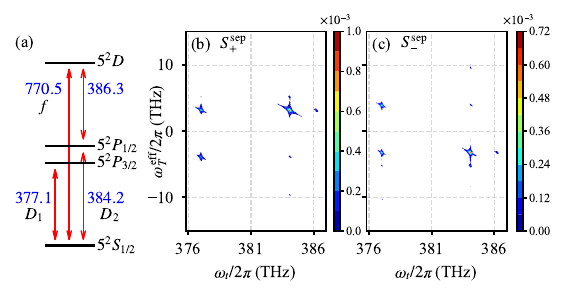}\caption{Energy level diagram and the 2D spectra of $^{87}\textrm{Rb}$. (a)
The energy levels of an individual $^{87}\textrm{Rb}$ atom. All numerical
labels in the diagram are in units of $\text{THz}$. (b) and (c) show
the separated 2D spectra, $S_{+}^{\mathrm{sep}}(\tau,\omega_{T}^{\mathrm{eff}},\omega_{t};\omega_{\mathrm{RF}})$
and $S_{-}^{\mathrm{sep}}(\tau,\omega_{T}^{\mathrm{eff}},\omega_{t};\omega_{\mathrm{RF}})$,
respectively under the rotating frames at $787\text{nm}$. \label{fig:The-absolute-spectra}}
\end{figure}

The probe pulse after interaction with the sample is spectrally resolved using a spectrometer equipped with a charge-coupled device (CCD) camera.
These pulses have a bandwidth (FWHM) of 11.9THz and spectrally cover
the Rb $D_{1}$ transition ($377.1\text{THz}$, $795\text{nm}$),
$D_{2}$ transition ($384.2\text{THz}$, $780\text{nm}$) and the
excited state $5^{2}P_{3/2}\leftrightarrow5^{2}D$ transition ($386.3\text{THz}$,
$776\text{nm}$) as shown in Fig.~\ref{fig:The-absolute-spectra}(a).
Experiments are conducted under three rotating\nobreakdash-frame
frequencies: $380.9\text{THz}$ $(787\text{nm}$), $379.5\,\text{THz}$
($790\text{nm}$), and $378.05\text{THz}$ ($793\text{nm}$).

We apply the four-step phase cycling scheme with phase settings: $\Phi_{1}$,
$\Phi_{2}$, $\Phi_{3}$, and $\Phi_{4}$. The separated spectra $S_{+}^{\mathrm{sep}}(\tau,\omega_{T}^{\mathrm{eff}},\omega_{t};\omega_{\mathrm{RF}})$
and $S_{-}^{\mathrm{sep}}(\tau,\omega_{T}^{\mathrm{eff}},\omega_{t};\omega_{\mathrm{RF}})$
of $^{87}\textrm{Rb}$ atoms with different combination of measured
data are shown in Figs.~\ref{fig:The-absolute-spectra}(b) and (c),
with the rotating frames $\omega_{\mathrm{RF}}/2\pi=380.9\text{THz}$
($787\text{nm}$). The emission frequencies of the signal are $\omega_{t}/2\pi=377.1\text{THz}$,
$384.2\text{THz}$ and $386.3\text{THz}$, corresponding to the frequencies
of $D_{1}$ transition, $D_{2}$ transition and an excited state transition
of an individual $^{87}\mathrm{Rb}$ atom, respectively.

\begin{figure*}[!t]
\centering{}
\includegraphics[width=\textwidth]{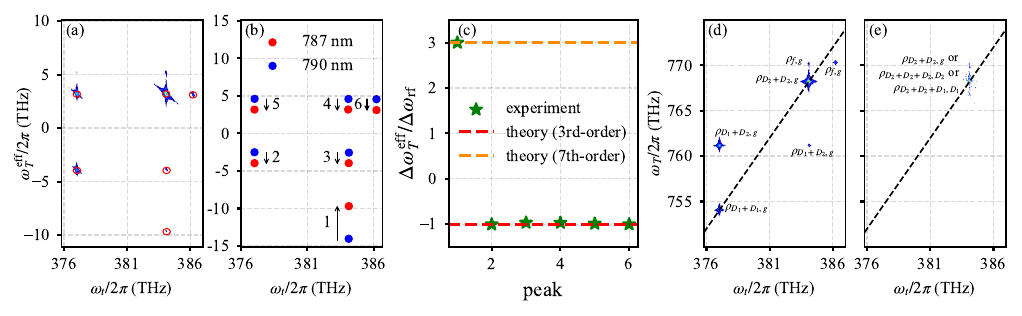}
\caption
{Results of the frame-shift tracking algorithm and the separated higher-order
spectra. (a) Red circles show the center peak centers of the signals,
extracted via two-dimensional Gaussian fitting from the $S_{+}^{\mathrm{sep}}(\tau,\omega_{T}^{\mathrm{eff}},\omega_{t};\omega_{\mathrm{RF}})$
spectrum at $787\text{nm}$ in the rotating frame. The numbers label
the peak indices. (b) Signal peaks in the rotating frame at $787\text{nm}$
(red points) and $790\text{nm}$ (blue points). The arrows indicate
the peak movements. (c) Shift rates $\Delta\omega_{T}^{\text{eff}}/\Delta\omega_{\mathrm{RF}}$
for each peak. Green pentagrams represent the signal peaks from (b),
with the horizontal axis indices corresponding to those in (a). The extracted shift rates for peaks 1-6 are 3.000, -1.006, -0.968, -0.970, -0.990, and -1.004, respectively. The
red and orange dashed lines denote the theoretical values for the
3rd- and 7th-order signals, respectively. (d) and (e) present the
frequency-retrieved 2Q spectra of the 3rd- and 7th-order signals extracted
from the $S_{+}^{\mathrm{sep}}(\tau,\omega_{T}^{\mathrm{eff}},\omega_{t};\omega_{\mathrm{RF}})$
spectrum, respectively. The black dotted line shows $\omega_{T}=2\omega_{t}$.
\label{fig:Signal-center-peaks}}
\end{figure*}

We first focus on the signal isolation for the spectrum $S_{+}^{\mathrm{sep}}(\tau,\omega_{T}^{\mathrm{eff}},\omega_{t})$
in the following discussions. With the Gaussian fitting in Eq. (\ref{eq:6}),
we identify 6 peaks, illustrated with the red circles in Fig.~\ref{fig:Signal-center-peaks}(a),
show the fitting results for the spectra $S_{+}^{\mathrm{sep}}(\tau,\omega_{T}^{\mathrm{eff}},\omega_{t})$
with the rotating frame set at $787\text{nm}$. With the same method,
we also find the peak centers $(\omega_{x},\omega_{y})$ for the rotating
frame at $790\text{nm}$, illustrated as the blue points in Fig.~\ref{fig:Signal-center-peaks}(b),
along with the peak centers for $787\mathrm{nm}$ as red points. Fig.~\ref{fig:Signal-center-peaks}(b)
clearly illustrates the different shift rates $\Delta\omega_{T}^{\text{eff}}/\Delta\omega_{\mathrm{RF}}$
for peak centers.

Then we apply the Hungarian algorithm to determine the shift rates
$\Delta\omega_{T}^{\text{eff}}/\Delta\omega_{\mathrm{RF}}$ for all
the 6 peaks with the results shown as blue pentagrams in Fig.~\ref{fig:Signal-center-peaks}(c),
where one peak has different rate $\Delta\omega_{T}^{\text{eff}}/\Delta\omega_{\mathrm{RF}}=3$
from other 5 peaks cluster around a common value of $-0.99 \pm 0.03$.
With such differences, we will isolate the signal by introducing a mask
functions for the $i$-th peak as 
\begin{equation}
\textrm{MF}_{i}(\omega_{t},\omega_{T}^{\mathrm{eff}})=\begin{cases}
1 & \textrm{if}\:d<d_{0},\\
2/(1+\gamma\exp[d-d_{0}]) & \textrm{if}\:d\geq d_{0},
\end{cases}
\end{equation}
where the index distance $d$ of the point ($\omega_{t},\omega_{T}^{\mathrm{eff}}$)
from $i$-th signal center $(\omega_{x}^{(i)},\omega_{y}^{(i)})$.
And $d_{0}=20$ is a spatial threshold value and $\gamma=0.05$ controls
the smoothness of the mask function. Then, the mask function for the
same shift rate $\Delta\omega_{T}^{\text{eff}}/\Delta\omega_{\mathrm{RF}}$
is generated by the summation as $\mathrm{MF}(\omega_{t},\omega_{T}^{\mathrm{eff}};-1)=\sum_{i=2}^{6}\textrm{MF}_{i}$
and $\mathrm{MF}(\omega_{t},\omega_{T}^{\mathrm{eff}};3)=\textrm{MF}_{1}$.
And the different order signals are obtained by multiplying the original
spectrum $S_{+}^{\mathrm{sep}}(\tau,\omega_{T}^{\mathrm{eff}},\omega_{t})$
with the two mask functions $\mathrm{WF}(\omega_{t},\omega_{T}^{\mathrm{eff}};-1)$
and $\mathrm{WF}(\omega_{t},\omega_{T}^{\mathrm{eff}};3)$. And the
frequencies for the spectra are retrieved by adding $\omega_{t}+\omega_{\mathrm{RF}}$
to $\omega_{T}^{\text{eff}}$ for the shift rate $\Delta\omega_{T}^{\text{eff}}/\Delta\omega_{\mathrm{RF}}=-1$
and adding $\omega_{t}-3\omega_{\mathrm{RF}}$ to $\omega_{T}^{\text{eff}}$
for the shift rate $\Delta\omega_{T}^{\text{eff}}/\Delta\omega_{\mathrm{RF}}=3$.
Figure~\ref{fig:Signal-center-peaks}(d) shows the frequency-retrieved
2Q spectra for the 3rd-order signal ($\Delta\omega_{T}^{\text{eff}}/\Delta\omega_{\mathrm{RF}}=-1$),
while Fig.~\ref{fig:Signal-center-peaks}(e) shows the 2Q spectra
for the 7th-order signal $(\ensuremath{\Delta\omega_{T}^{\text{eff}}/\Delta\omega_{\mathrm{RF}}=3})$. 

The 2Q spectrum of $^{87}\text{Rb}$ in the panel (d) exhibit peak
locations consistent with those reported in other works \cite{Gao2016OptLett,Yan2022,Cai2024}.
In particular, in Fig.~\ref{fig:Signal-center-peaks}(d), the peaks
at $(\omega_{t}/2\pi=384.2\text{THz},\omega_{T}/2\pi=770.5\text{THz})$
and $(\omega_{t}/2\pi=386.3\text{THz},\omega_{T}/2\pi=770.5\text{THz})$
correspond to the doubly excited state of an individual $^{87}\text{Rb}$
atom. The other four peaks correspond to collective resonances resulting
from dipole--dipole interactions in $D_{1}+D_{1}$, $D_{1}+D_{2}$,
and $D_{2}+D_{2}$ transitions between two $^{87}\text{Rb}$ atoms.
In Fig.~\ref{fig:Signal-center-peaks}(e), the peak at $(384.2\text{THz},768.4\text{THz})$
arises from collective resonances generated by the dipole-dipole interaction
among the multiple $^{87}\text{Rb}$ atoms. Signals in Fig. \ref{fig:Signal-center-peaks}(d)
are of the third order, with the phase\nobreakdash-matching condition
$\mathbf{k}_{s}=\mathbf{k}_{2}+\mathbf{k}_{1}-\mathbf{k}_{1'}$. For
the two peaks at $(\omega_{t}/2\pi=384.2\text{THz},\omega_{T}/2\pi=770.5\text{THz})$
and $(\omega_{t}/2\pi=386.3\text{THz},\omega_{T}/2\pi=770.5\text{THz})$,
the field pair $\mathbf{k}_{2}+\mathbf{k}_{1}$ first creates a coherence
$\rho_{f,g}$ during the time interval $T$, and the subsequent action
of $-\mathbf{k}_{1'}$ generates either $\rho_{D_{2},g}$ or $\rho_{f,D_{2}}$
during the detection time $t$. Here the state $\left|f\right\rangle $
corresponds to the $5^{2}S_{1/2}\leftrightarrow5^{2}D$ transition,
prepared by the probe and the first pump pulse. The two peaks at $(\omega_{t}/2\pi=377.1\text{THz},\omega_{T}/2\pi=761.3\text{THz})$
and $(\omega_{t}/2\pi=384.2\text{THz},\omega_{T}/2\pi=761.3\text{THz})$
originate from a coherence $\rho_{D_{1}+D_{2},g}$ prepared by $\mathbf{k}_{2}+\mathbf{k}_{1}$
during the time interval $T$. The final interaction with $-\mathbf{k}_{1'}$
yields either $\rho_{D_{1}+D_{2},D_{2}}$ (or effectively $\rho_{D_{1},g}$
) or $\rho_{D_{1}+D_{2},D_{1}}$ (or effectively $\rho_{D_{2},g}$
), accounting for the two observed spectral positions. The remaining
two peaks follow analogous pathways. And the peak in Fig. \ref{fig:Signal-center-peaks}(e)
corresponds the 7th-order signal (see line 2 and 3 of Table \ref{table1_em}), which was seldomly isolated in previous studies.

\begin{figure}[t]
\centering
\centering{}\includegraphics[width=8cm]{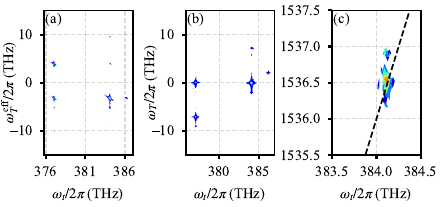}\caption{The $S_{-}^{\mathrm{sep}}(\tau,\omega_{T}^{\mathrm{eff}},\omega_{t};\omega_{\mathrm{RF}})$
spectrum and the separated higher-order contributions. (a) The $S_{-}^{\mathrm{sep}}(\tau,\omega_{T}^{\mathrm{eff}},\omega_{t};\omega_{\mathrm{RF}})$
spectrum at $787\text{nm}$ in the rotating frame, identical to Figs.~\ref{fig:The-absolute-spectra}(c).
(b) The retrieved 0Q spectrum of the 3rd-order signal, obtained using
the same retrieval procedure as described in Fig.~\ref{fig:Signal-center-peaks}. (c) The
recovered 4Q spectrum of the 7th-order signal.\label{fig:S-spectrum}}
\end{figure}

The separation procedure for the $S_{-}^{\mathrm{sep}}(\tau,\omega_{T}^{\mathrm{eff}},\omega_{t};\omega_{\mathrm{RF}})$
spectra is identical to that applied to $S_{+}^{\mathrm{sep}}(\tau,\omega_{T}^{\mathrm{eff}},\omega_{t};\omega_{\mathrm{RF}})$
. Starting from the $S_{-}^{\mathrm{sep}}(\tau,\omega_{T}^{\mathrm{eff}},\omega_{t};\omega_{\mathrm{RF}})$
spectrum, as shown in Fig. \ref{fig:S-spectrum}(a), 3rd-order and
7th-order contributions are identified in the same manner by identifying
different shift rates. By further applying the corresponding frequency-retrieval
conditions, the separated 3rd-order 0Q coherence signal and the 7th-order
4Q coherence signal are obtained, as shown in Fig. \ref{fig:S-spectrum}(b)
and Fig. \ref{fig:S-spectrum}(c), respectively. The observed 7th-order 0Q contribution corresponding to line 6 in Table. \ref{table1_em} is successfully isolated within the 0Q spectrum, as shown in Fig. \ref{fig:S-spectrum}(b).

For other potential signal peaks arising from multi-atom interactions, such as the $\rho_{3,1}$ coherence corresponding to line 4 in Table. \ref{table1_em}, no distinct spectral features were experimentally resolved. This is likely due to weak interaction strengths or minimal energy-level shifts, and these contributions are therefore presented primarily as theoretically possible pathways.

With the $^{87}\mathrm{Rb}$ vapor system, we demonstrate the experimental
isolation of the 7th-order nonlinear signal with the computation-assisted method. And the current method, tailored for multidimensional
spectroscopy, uses four phase settings along with experimental acquisitions
with three rotating frames, and substantially reduces experimental
difficulties with complex phase cycling.

\section{Conclusion}
We have demonstrated an order-selective strategy
for high-order nonlinear spectroscopy that combines rotating-frame
acquisition with frame-shift tracking analysis of frame-dependent spectral
shifts. In this framework, the rotating frame is used not only for
sampling efficiency but also as a controlled modulation axis that
encodes perturbative order through distinct peak-shift slopes. This
enables systematic separation of overlapping nonlinear contributions
after standard phase-cycling processing.

Using two-dimensional spectroscopy of rubidium vapor, we experimentally
isolate a seventh-order response from coexisting third-order backgrounds,
thereby establishing direct access to beyond-third-order coherence
dynamics in a practical excitation regime. This approach is scalable to higher-order signals, provided that the corresponding pathways exhibit resolvable frame-dependent spectral shifts and sufficient signal-to-noise ratio.

\section{Back matter}

\begin{backmatter}
\bmsection{Funding}
This work is supported
by the Quantum Science and Technology-National Science and Technology
Major Project (Grant No. 2023ZD0300700), and the National Natural
Science Foundation of China (Grant Nos. U2230203, U2330401, 12088101).

\bmsection{Acknowledgment}
We thank Dr.~Mao-Rui Cai for helpful discussions.

\bmsection{Disclosures}
The authors declare no conflicts of interest.

\bmsection{Data Availability Statement}
Data underlying the results presented in this paper are not publicly available at this time but may be obtained from the authors upon reasonable request.

\end{backmatter}

\bibliography{ref}

\end{document}